\documentclass[onecolumn,manuscript,showpacs]{revtex4}

\usepackage{graphicx}%
\usepackage{dcolumn}
\usepackage{amsmath}
\usepackage[dvips]{epsfig}      

\makeatletter
\def\btt#1{\texttt{\@backslashchar#1}}%
\DeclareRobustCommand\bblash{\btt{\@backslashchar}}%
\makeatother

\begin{document}

\title{ Electronic States and Magnetism of Mn Impurities and Dimers
       in Narrow-Gap and Wide-Gap III-V Semiconductors}

\author{W. H. Wang$^{1,2}$, Liang-Jian Zou$^{1,\footnote{Correspondence author, 
Electronic mail: zou@theory.issp.ac.cn}}$ and Y. Q. Wang$^1$}
\affiliation{\it
 $^1$Key Laboratory of Materials Physics, Institute of Solid State Physics,
Chinese Academy of Sciences, P. O. Box 1129, Hefei 230031, China}
\affiliation{\it $^2$Graduate School of the Chinese Academy of Sciences}

\date{Nov 22, 2004}

\begin{abstract}
   Electronic states and magnetic properties of single $Mn$ impurity and  
dimer doped in narrow-gap and wide-gap $III$-$V$ semiconductors have been 
studied systematically. It has been found that in the ground state for single 
$Mn$ impurity, $Mn$-$As(N)$ complex is antiferromagnetic (AFM) coupling when
$p$-$d$ hybridization $V_{pd}$ is large and both the hole level $E_{v}$ and
the impurity level $E_{d}$ are close to the midgap; or very weak ferromagnetic 
(FM) when $V_{pd}$ is small and both $E_{v}$ and $E_d$ are deep in the valence 
band. In $Mn$ dimer situation, the $Mn$ spins are AFM coupling for half-filled 
or full-filled $p$ orbits; on the contrast, the Mn spins are 
double-exchange-like
FM coupling for any $p$-orbits away from half-filling. We propose the strong 
{\it p-d} hybridized double exchange mechanism is responsible for the FM order 
in diluted $III$-$V$ semiconductors.
\end{abstract}

\pacs{75.50.Pp, 75.10.-b, 75.30.Hx}
\maketitle

  Diluted magnetic semiconductors (DMS), such as $Ga_{1-x}Mn_xAs$ and
$Ga_{1-x}Mn_xN$, with ferromagnetic (FM) Curie temperature ($T_c$)
as high as $110K$ in $Ga_{1-x}Mn_xAs$, or even as room temperature 
in $Ga_{1-x}Mn_xN$ have attracted much
attention, since these properties provide perspective applications
in the fabrication of spintronics devices as well as in quantum
computers [1-5]. The unusual FM order in $Mn$- or $Cr$-doped $III$-$V$ 
semiconductors with such low concentrated magnetic ions and 
the interesting magnetotransport properties have also raised
many fundamental problems, e.g. the electronic states of $3d$
impurities and the origin of the FM long-range order (LRO) [2-4]. 
Many experiments have established the fact that local magnetic moments 
{\bf S} and hole states are simultaneously introduced in DMS as $Ga$ 
atoms are substituted by $Mn$ ions. An elaborated relation between 
doped concentration and $T_c$ in Ref.[1]
demonstrated the important role of mobile hole carriers in the 
formation of FM LRO in DMS, leading to the hypothesis that the FM 
coupling between $Mn$ spins are mediated through these delocalized holes.

Several microscopic mechanisms have been proposed to explain the origin 
of the carrier-induced FM LRO in DMS. In the presence of local magnetic ions 
and mobile carriers, the RKKY interaction seems to 
be a natural candidate [6]. Magnetic properties of $Ga_{1-x}Mn_xAs$ 
were interpreted in the RKKY scenario in the mean field (MF) approximation.
Although the theory successfully explained the dependence 
of Curie temperature $T_c$ and spontaneous magnetization on doping 
concentration measured in experiments in some doping range [7],
there are still some important issues to be taken into account.  
The RKKY model validates only when isolated local spins merge
in the sea of free carriers and the carrier bandwidth $D$ is much 
greater than the spin-hole exchange constant $J$. However in DMS, the hole 
concentration $n_{h}$ is much less than local spins concentration  
$N_{s}$, and the $Mn$-$As(N)$ spin-hole exchange constant, $J$$\sim$1eV, 
is much larger than the Fermi energy $E_F$ of the hole carriers, about 
0.3 eV [8]. Furthermore, the $Mn$ concentration is so small that the 
nearest neighbour number $z$ is less than that of the conventional 
$3$-dimensional systems, thus it is expected that there will be
a large discrepancy between the MF result and realistic situation. 
Therefore the MF RKKY 
theory might be not accurate for describing the FM ordering in DMS.

The double exchange model which is responsible for the FM order in 
doped perovskite manganites was also suggested for the FM coupling 
of the distant $Mn$ spins in DMS [8, 9]. However, the spatial separation 
of $Mn$ spins in DMS is much further than dense $Mn$ spins in 
manganites, and
the hole wavefunction extends over tens of lattice sites in doped $GaAs$ 
and the valence fluctuation in Mn 3d orbits is small, dissimilar to the 
mixed valence of $3d^{3}$ and $3d^{4}$ in doped manganites. 
At present, whether or not the conventional double exchange mechanism 
works for the nature of FM in DMS does not get strong support 
experimentally and theoretically. 
The polaronic mediated FM mechanism [10] and some other theories are also 
proposed to interpret the origin of FM LRO, which shows that more efforts 
are needed for the microscopic origin of the FM order in DMS.

One of the central problems in the debate on the microscopic origin 
of FM LRO in DMS is that the single  
impurity electronic states and the spin interaction of two magnetic 
ions, especially its evolution with various parameters of DMS, 
are not well understood. Single $Mn$ impurity in $III$-$V$ and $IV$ 
semiconductors had been studied by Zunger et al. [11, 12] based on local 
density approximation and unconstrained MF approximation, substitutional 
impurity state properties of different transition-metal atoms 
in $III$-$V$ semiconductors had also been studied in Ref.[13], 
the dependence of the impurity energy level on the nuclei charge
of the transition metal atom had been known. 
The FM ground state (GS) in Mn doped DMS has also been 
confirmed by a lot of authors based on the first-principle electronic 
structure calculation, for example, see Ref.[14] and [15]. 
However these numerical studies are hard to clarify the microscopic 
processes of interacting $Mn$ spins and the evolution of 
electronic states and magnetic properties of the $Mn$ atoms and the 
host atoms in DMS with the doping concentration, the $p$-$d$ 
hybridization strength, the energy gap of the host $III$-$V$ semiconductors, 
the energy levels of the 3d impurities and the $4p$ holes, etc.. 
How these factors affect the FM ordering is crucial for our 
understanding of the unusual magnetism and transport in DMS.

The aim of this {\it Communication} is to elucidate the electronic 
states, the magnetic properties of single $Mn$-$As(N)$ complex, and the 
spin coupling between $Mn$ spins for various parameters. 
In the rest of this {\it Communication}, after describing the model 
Hamiltonian, we first study the GS of single $Mn$-$As(N)$ for various 
electron configurations, determining the parameter range of the 
antiferromagnetic (AFM) coupling in the $Mn$-$As(N)$ complex; 
then we study the $Mn$-$Mn$ spin couplings in the GS of two $Mn$-$As$(N) 
complexes,i.e. the Mn dimer, 
and show that the double-exchange-like mechanism is responsible for the FM 
order when the $p$-orbits of $As(N)$ is away from the half-filling, 
suggesting the intrinsic compensation of antisite defect or the inhomogeneity 
of hole distribution plays crucial roles.

For $Mn$ doped $III$-$V$ semiconductors with zinc-blende structure, such as 
$In$, $Ga$, or $Al$ for $III$ group elements and $As$ or $N$ for $V$ group 
elements, the five valence electrons of $As(N)$ occupies the four $sp^3$ 
dangling bonds with symmetry $A_1$ and $T_2$, the two $4s$ electrons of 
$Mn$ ion covalently 
couple to the $A_1$ orbit of the $As$ or $N$ dangling bonds. When 
the divalent $Mn$ impurity substitutes the trivalent $In$, $Ga$ or $Al$ as an
effective mass acceptor, we now have a basic interaction scenario in DMS from 
various early experimental and theoretical studies: $Mn$ ion contributes a 
spin and a hole, the spin localizes in $Mn$ site, while the hole is bound 
to $As$ or $N$ site with 
an extension radius of $r_{s}$ much larger than the lattice constant; 
the $3d$ electrons in $Mn$ impurity interacts with the holes in the $p$ 
orbits of surrounding $As$ or $N$ sites through hybridization $V_{pd}$, 
forming the $Mn$-$As (N)$ complex. Under $T_{d}$ symmetry environment of 
the zinc-blende 
$III$-$V$ semiconductors, the hole has equi-possibility to occupy the T$_{2}$ 
orbit in one of the four nearest-neighbour $As$ or $N$ sites around $Mn$
ion, the hole is $4$-fold degenerate; and in the $T_{d}$ crystalline field,
the $3d$ orbits of $Mn$ ions split into lower energy $E_{g}-$ like $d$ orbits 
with pure atomic character and not coupling to $As$ or $N$, and higher energy 
$T_{2g}$-like orbits coupling to the $T_{2}$ $p$ orbit of $As$ or $N$ [13]. 
The $Mn$ $T_{2g}$ electron which hybridizes with the $As$ or $N$ hole 
is thus $3$-fold degenerate.
Due to the strong Hund's rule coupling, the hybridized $3d$ electron still 
couple with the local spin $\bf{S}$ ($S=2$) of $Mn$ $3d^4$ configuration 
via Hund's rule 
coupling $J_{H}$. The on-site Coulomb interactions between $3d$ electrons 
and between holes, $U_{d}$, and $U_{v}$, are also taken into account.

    With the increase of doped $Mn$ concentration, the wavefunctions of 
the holes centered at $As$ or $N$ atom with large radius $r_{s}$ start to 
overlap with each other, the hopping integral between holes, $V_{h}$, 
crucially depends on the extension radius $r_{s}$ and the hole density 
$n_{h}$. The Hamiltonian modeling the preceding physics 
in DMS thus reads:
\begin{eqnarray}
\hat{H}&=&\sum_{i}\hat{H_{0}}(i)+\hat{H_{1}}
\end{eqnarray}
\begin{eqnarray}
  \hat{H_{0}}(i)&=&\sum\limits_{\sigma} (E_{d}\hat{d}^{\dagger}_{i\sigma}\hat{d
      }_{i\sigma} + E_{v}\hat{c}^{\dagger}_{i\sigma}\hat{c}_{i\sigma}
+\frac{U_{d}}{2} n^d_{i\sigma}n^d_{i\overline{\sigma}}
+\frac{U_{v}}{2}n^c_{i\sigma}n^c_{i\overline{\sigma}}) \nonumber \\
& &-\frac{J_{H}}{2}\sum\limits_{\mu\nu}{\bf{S_i}}\cdot\hat{d}^{\dagger}_{i\mu}
  {\bf{\sigma_{i\mu\nu}}}{\hat{d}_{i\nu}}
+\sum\limits_{\sigma}V_{pd}(\hat{d}^{\dagger}_{i\sigma}\hat{c}_{i\sigma} + 
\hat{c}^{\dagger}_{i\sigma}\hat{d}_{i\sigma})
\end{eqnarray}
\begin{eqnarray}
\hat{H_{1}}&=&\sum\limits_{<ij>\sigma}V_h(\hat{c}^{\dagger}_{i\sigma}
  \hat{c}_{j\sigma}+\hat{c}^{\dagger}_{j\sigma}\hat{c}_{i\sigma})
\end{eqnarray}
where $H_{0}$(i) and $H_1$ represent the interactions in the $i$th 
$Mn$-$As(N)$ complex and the inter-complexes hopping, respectively;
$E_{d}$ and $E_{v}$ are the bare $d$-electron and $p$-orbit
energy levels, respectively; $\hat{d}^{\dagger}_{i\sigma}$ and
$\hat{c}^{\dagger}_{i\sigma}$ denote the creation operators of the $d$ 
electron and $p$ electrons with spin $\sigma$ in the $i$th $Mn-As(N)$ 
complex; we separate $Mn$ $3d$ electrons into an inert local spin 
${\bf{S_i}}$ of $3d^4$ configuration and a $3d$ electron which hybridizes 
with $4p$ orbit, the $3d$ electron couples with the local 
spin via strong Hund coupling;
$n^d_{i\sigma}$ and $n^c_{i\sigma}$ denotes the electron occupation 
number of $d$ state and $p$ state with spin $\sigma$, and 
n=$\sum_{\sigma}n^c_{\sigma}$+$n^d_{\sigma}$ the total electron number; 
here $U_{v}$ is usually much smaller than $U_{d}$ [13];
unlike many other authors, we do not write our Hamiltonian as the simple 
Kondo lattice model, since we recognize that 
there exists strong $pd$ hybridization between $Mn$ $3d$ orbits and $As$ 
or $N$ $p$ orbits, the $3d$ electron occupation varies with the $pd$ 
hybridization, and 
the hole in the $p$ orbit extends over a few of lattice constants, thus the 
Kondo lattice model is not proper to describe the physics in DMS.
In this paper, we take the chemical potential $\mu$ located at the 
middle of the band gap, and the $p$- and $d$-levels $E_{v}$ and $E_{d}$ are 
taken with respect to $\mu$. For clarity we neglect the degeneracy of $Mn$ 
$3d$ electrons and $p$ holes in the present study.

In the following we first present the electronic states and the evolution 
of single $Mn$ impurity with the interaction parameters, and then the 
effective coupling between $Mn$ spins of two $Mn$-$As$ complexes in the 
DMS background.
\begin{figure}[tp]
\vglue -0.6cm
\scalebox{1.150}[1.15]{\epsfig{figure=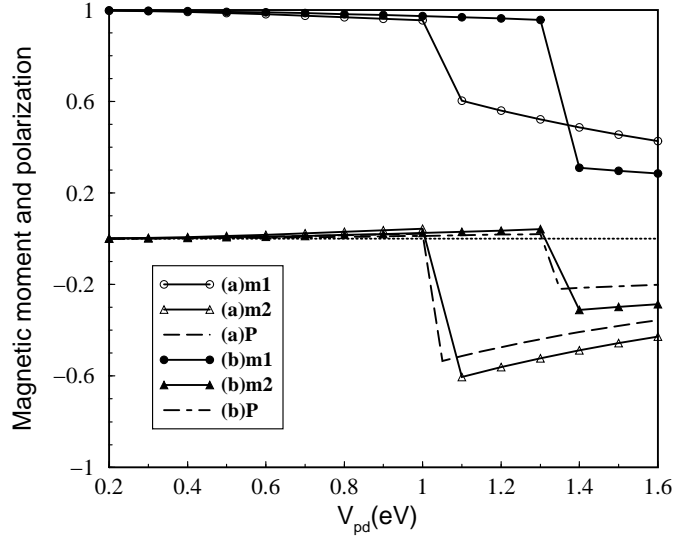,width=7.0cm,angle=270.0}}
\caption{Hybridization dependence of magnetic moment $m_1$, $m_2$ of the $d$,  
$p$ electrons and spin polarization $P$ of $p$ electrons in the GS 
of $Mn-As(N)$ complex for
(a) $GaAs$:$Mn$, $E_v$=-0.95 eV, $E_d$=-1.2 eV;
(b) $GaN$:$Mn$, $E_v$=-1.8 eV, $E_d$=-0.5 eV. The other parameters:
$U_d$=4.0 eV, $U_v$=0.35 eV, $J_H$=1.0 eV.}
\label{fig:fig1}
\end{figure}
\\

\noindent{\bf \it{A}. Single Mn Impurity }

   We first study the electronic states of $Mn$ and its ligand $As$ or $N$ 
atoms in the diluted limit, which describes the case that the distance of 
$Mn$ ions is 
so far that the interaction between $Mn$ impurities can be neglected, and 
no summation over lattice is needed in the Eq.$(1)$-$(3)$. 
The local spin $\bf{S}$ is assumed to be aligned in the z axis, 
the electronic states and the magnetic properties of Mn-As complex 
are easily obtained by the exact diagonalization of H$_{0}$ for various 
physical parameter and electron configurations.
As a contrast to the local moments formation in the Anderson impurity model, 
the $p$-electron is AFM polarized as the $p$-$d$ hybridization $V_{pd}$ is 
stronger than a critical value $V_c$, as seen in Fig. 1. When $V_{pd}$ is 
small, $V_{pd}<V_c$, 
we find that in the GS the $p$ orbit is almost full filled and 
the electrons in each $p$ orbit are positive polarized weakly. 
The $p$ and $d$ electron occupations are almost fixed for $V_{pd} < V_c$. 
The maximized polarized moment of each $As$ or $N$ are about $0.026\mu_B$ 
for $GaAs$:$Mn$ and $0.042\mu_B$ for $GaN$:$Mn$ as $V_{pd}\rightarrow{V_c}$. 
A small fraction of spin-down $p$ electron transfers to $3d$ orbit, 
which leads to the weak positive polarization of the $p$ orbit and 
$S\approx\frac{5}{2}$ at $Mn$ site. 
As $V_{pd}$ increases to $V_c$ $\approx{0.88 eV}$ for $GaAs$:$Mn$ or 
$\approx{1.33 eV}$
for $GaN$:$Mn$, the holes strongly hybridize with the $3d$ electron, 
a large fraction of $d$ electron transfers to the $p$ orbit, leading to 
the AFM alignment of the $p$ electron in respect to the $d$ electron.
This result is in agreement with the magnetic circular dichroism experiment 
[16]. The corresponding polarization of the $As$ or $N$ atoms around $Mn$ 
impurity, is also shown in Fig.1. 
The maximum polarized magnetic moment of each $As(N)$ is about 
$-0.58\mu_B$ for $GaAs$:$Mn$ and $-0.32\mu_B$ for $GaN$:$Mn$. 
Our result for $GaN$:$Mn$ is in good agreement with that obtained by the 
first-principles electronic structure calculation in Ref. [17]
for Mn-N cluster, but considerable larger than that in periodic systems [18]. 
While, considering the four equivalent $As$ or $N$ atoms around the $Mn$ ion, 
we expect the polarized moment of each $As$ or $N$ atom is smaller than this 
value.

   The dependence of the magnetic moment and the polarization of the $p$ 
orbits on the $3d$ energy level $E_d$ is very similar to that on the 
hybridization $V_{pd}$ in Fig.1. For deep $3d$ energy level $E_d$, the $p$ 
electrons in $p$ orbits are positively polarized and very weak; there 
also exists a critical value $E^{c}_{d}$
that when $E_d$ is shallower than $E^{c}_{d}$, the electrons in the $p$ orbits 
are AFM polarized, the polarization of the $p$ orbit is about 
$-40\%$ for $GaAs$:$Mn$ or $-20\%$ for $GaN$:$Mn$.
Since the hole is located around the top of valence band, the closer the 
$E_{d}$ is to $E_{v}$, the larger the $p$-$d$ hybridization is, hence 
more strong polarization of
the $p$ electron. Too deep or too shallow $E_d$ level is not favorable of 
the formation of the AFM $Mn$-$As(N)$ complex.
In the GS for $V_{pd}=1.0eV$, the energy difference between FM and AFM 
configurations of $Mn$-$As(N)$ complex is $-0.93 eV$ for $GaAs$:$Mn$ and 
$-1.26eV$ for $GaN$:$Mn$. The former agrees with the photoemission data 
obtained by J. Okabayashi et al. [19].
\begin{figure}[tp]
\vglue -0.6cm
\scalebox{1.150}[1.15]{\epsfig{figure=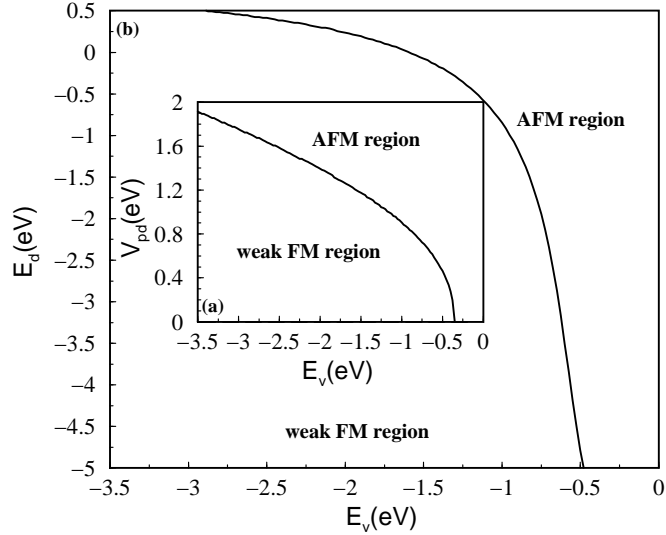,width=7.0cm,angle=270.0}}
\caption{The phase diagram of $Mn$-$As(N)$ complex (a) $V_{pd}$ {\it vs} $E_v$, 
$E_d$=-0.5 eV and  (b) $E_{d}$ vs $E_v$, $V_{pd}$=1.0 eV. The other 
parameters are the same in Fig.1.}
\label{fig:fig2}
\end{figure}

The magnetic phase diagrams of single $Mn$-$As(N)$ complex on $V_{pd}$ 
{\it vs} $E_d$ and {\it vs} $E_v$ are shown in Fig.2a and Fig.2b. The common 
character of the phase diagrams is there exist very weak FM and AFM polarized 
regions. We find when both $E_d$ and $E_v$ are very deep, the electrons 
in the $p$ orbits are not polarized, i.e. the paramagnetic region, this 
region is not shown in Fig.3; as $E_d$ and $E_v$ are lifted,
the $p$ electrons become weak FM polarization to the $Mn$ spin;
and for large enough $V_{pd}$ and shallow $E_d$ or $E_v$, the $p$ electrons in 
$As$ or $N$ site are AFM polarized. 
In fact, the physical parameters in realistic DMS $GaAs$:$Mn$ and $GaN$:$Mn$ 
fall into the AFM region, this provides 
such a possibility that the $Mn$ spins interact through 
AFM polarized $As$ or $N$ ligands to form FM correlation. 
As we showed in the following, the $Mn$ spins establish double-exchange-like 
FM correlation via the polarized $p$-$d$ hybridized band. 
\\

\noindent{\bf \it{B}. $Mn$ Dimers }

Next we consider two $Mn$-$As(N)$ complexes, and study 
the GS magnetic configuration and the spin coupling between $Mn$ ions.
With the increasing of $Mn$ doping density and hole concentration,
the separation between holes becomes smaller and smaller, the wavefunctions of
these holes begin to overlap. The hopping integral between two overlapped holes 
at sites $\bf{R_i}$ and $\bf{R_j}$ is depicted by
$V_h(i,j)=<\psi_h{(\bf{R_i})}|h_0|\psi_h{(\bf{R_j})}>$,
here $\psi_h{(\bf{R_i})}$ is the hole wavefunction and $h_0$ is the 
Hamiltonian of single particles. We approximate $\psi_h$ with the hydrogen-like 
wave function with radius $r_s$ and effective mass $m^*$. In this situation, 
the summation of lattice in Eq.(1)-(3) runs over indices $1$ and $2$.
For simplify, the core spins of two $Mn$ ions are assumed semiclassical 
and the spin $\bf {S_2}$ deviates an angle of $\theta$ with respect to 
$\bf {S_1}$.

In the GS of $Mn$ dimer, the $Mn$ core spins are either strong FM or weak AFM 
coupling, depending on the electron filling the $p$-orbits, as shown in Fig.3. 
\begin{figure}[tp]
\vglue -0.6cm
\scalebox{1.150}[1.15]{\epsfig{figure=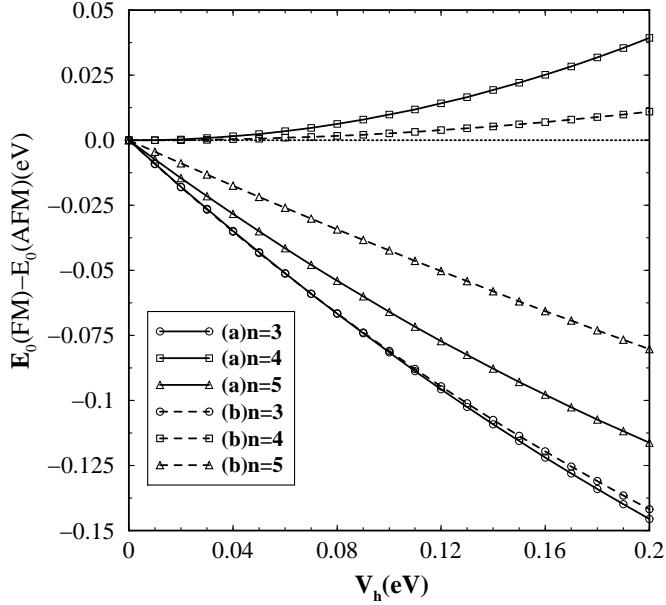,width=8.0cm,angle=270.0}}
\caption{The ground state energy difference between $Mn$-$Mn$ FM coupling 
and AFM coupling $vs$. $V_h$ in $(Mn-As)$ dimer for (a) $GaAs$:$Mn$ and (b) 
$GaN$:$Mn$,
$V_{pd}$= 1.0eV. The other parameters are the same in Fig.1.}
\label{fig:fig3}
\end{figure}
Comparing the GS energy of the $Mn$ dimer of FM configuration with that of 
AFM configuration, we find that $Mn$-$Mn$ AFM configuration is stable for
the half- or full-filled $p$-orbits in the two $As$ or $N$ sites, 
however, FM is more stable when the electron filling in the $As$ or $N$ 
$p$ orbit deviates from half- or full- filling;
and the more the hole number is, the stronger the $Mn$-$Mn$ FM coupling is.
The magnetic coupling strength of the $Mn$ dimer monotonically increases with 
the hopping integral $V_h$ and the AFM coupling strength is much less 
than the FM coupling strength, which can be clearly seen in Fig.3

With the substitution doping of $Mn$ to $Ga$, core spins and holes are 
introduced simultaneously in the semiconductor host. Formally the density
of $Mn$ spins equals to the hole concentration. It seems that the active 
$p$ orbits is always half-filled. For the half-filled $Mn$ dimer,
the opposite spin alignment of the different $p$ orbits, 
the AFM coupled $p$-$d$ hybridization, the strong Hund's coupling 
between the $3d$ electron and the local spin lead to AFM coupling of the $Mn$ 
local spins. However, considering the existence of numerous antisite $As$ 
and the intersite $Mn$ defects,
a significant fraction of holes are compensated by the electrons from 
these defects.
Thus the hole concentration is considerable less than the $Mn$ density, the
$n=5$ electron configuration in the $(Mn$-$As)$ dimer is the most probable in 
realistic DMS. In the GS with electron configuration away from 
half-filling, the hopping of holes and their AFM hybridization with $3d$ 
electrons lead to the FM coupling between $Mn$ 
$3d$ electrons, hence the FM coupling of $Mn$-$Mn$ spins, just as our 
prediction for $GaAs$:$Mn$ and $GaN$:$Mn$.

   To further understand the nature of FM LRO in DMS, we study the GS 
total energy on the azimuthal angle $\theta$ between two local spins for 
various electron filling, and the result is shown in Fig.4.
\begin{figure}[tp]
\vglue -0.6cm
\scalebox{1.150}[1.15]
{\epsfig{figure=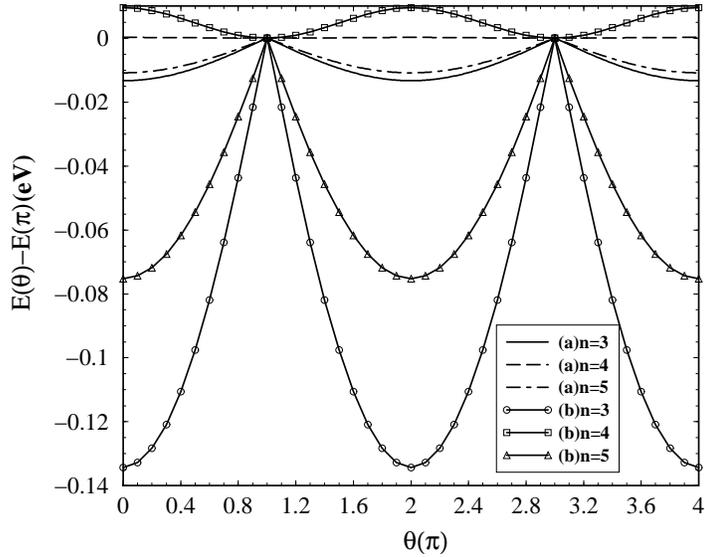,width=7.0cm,angle=270.0}}
\caption{ The $\theta$ dependence of ground-state energy difference 
in two ($Mn$-$As$) complexes for 
(a) $GaAs$:$Mn$, $V_h\approx0.015eV$ and (b) $GaN$:$Mn$, $V_h\approx$ 0.186 eV.
The other parameters are the same in Fig.3.}
\label{fig:fig4}
\end{figure}
The total energies $E(\theta)$ are measured relative to the $Mn$-$Mn$ AFM 
configuration. We notice that for the half-filling case, the energy 
difference exhibits $\cos\theta$ 
behavior, indicating that the $Mn$-$Mn$ coupling is Heisenberg-like AFM. 
In contrast, in the case deviating from half-filling, the GS energy difference 
can be well fitted by $|\cos\frac{\theta}{2}|$, and this character is very 
similar to doped manganites, which implies that the $Mn$-$Mn$ FM coupling is 
double-exchange-like. This outstanding behavior seems to conflict with 
what we fore-mentioned. In fact, in the present scenario of strong 
hybridization with $p$ orbit and strong Hund's coupling with core spin in the 
DMS, the $d$-electron form a narrow itinerant band, so the physical 
interaction of $Mn$ spins in proper doping DMS quite resembles double exchange 
interaction in mixed-valence manganites. 
The FM coupling energy of two $Mn$ spins in $GaAs$:$Mn$ is about $11$ meV, 
comparable with experimental Curie temperature $T_c$. In $GaN$:$Mn$ the FM 
coupling energy is about $75$ meV, 
it seems plausible to explain the origin of extremely high Curie temperature 
$T_c\approx 600-900K$ in wide-gap $GaN$:$Mn$.

  We find in AFM polarized $Mn$-$As$ and $Mn$-$N$, the polarized 
magnetic moment at $As$ site is larger than that at $N$ site, this occurs
both for single complexes and for dimers; on the contrary, the FM coupling 
energy of the $Mn$ spins in $GaAs$:$Mn$ is smaller than that in $GaN$:$Mn$. 
This arises from that the charge transfer of the spin-up $3d$ electron 
to the $p$-orbit in $Mn$-$As$ is less than that in $Mn$-$N$; however, 
the hopping integral of the holes in the latter is larger than the former, 
leading to more wider hybridized $d$-band in the latter and contributing 
more strong double-exchange FM coupling strength. 
From Fig.3 and Fig.4, one find that the more the hole number is, 
the stronger the $Mn$-$Mn$ magnetic coupling is, implying that 
the presence of more holes favors the FM in DMS. The observation of FM
ordering in DMS suggests that antisite As or intersite $Mn$ atoms may 
compensate a significant fraction of holes, leading to the observed
double-exchange-like FM coupling strength. Also, the possible inhomogeneous 
distribution of the hole density could lead to the FM phase, since 
even in the molecular-beam epitaxy grown DMS thin films inhomogeneous 
strain was found to widely exist in most of DMS samples, this favors
the inhomogeneous distribution of the holes. 
With the coexistence of both the $n$=$3$ and the $n$=$5$ electron 
configurations, lower magnetic energy and inhomogeneous distribution of 
FM coupling are expected in the GS.

As we mentioned in the preceding, the $Mn$ local spins in DMS is so distant in 
comparison with dense spins in doped manganites that one may question the 
role of the double exchange mechanism in DMS. The microscopic mechanism 
for the double exchange FM in Mn-doped DMS is depicted as follows: 
due to the strong hybridization of the $As(N)$ $p$ orbits with the 3d 
electrons, the system forms two hybridized narrow bands; the hybridized 
band with dominant $d$ orbital character couples to the $Mn$ core spins 
via strong Hund's rule coupling, thus the hopping of the mobile electrons 
between localized $Mn$ spins gives rise to the double-exchange-like FM 
coupling and leads to the FM order in DMS, similar to that in doped manganites.

    Compare the present double-exchange-like FM with the conventional 
double exchange FM in manganites, we find the valence fluctuation in the FM 
ordered DMS is significantly less than that in doped manganites. Due to 
the strong $p$-$d$ hybridization and the extended wavefunctions of the $p$ 
holes, the lifetimes of the mobile electrons staying around two local spins 
are nearly equal, therefore one would not expect strong valence fluctuation 
in Mn 3d orbits.
Meanwhile, the Curie temperature in DMS seems to be too
high in comparison with the FM critical temperature in manganites, this may
attribute to that the strong Jahn-Teller e-ph coupling and polaronic effect 
weaken the FM phase transition point in manganites. 

    In summary, we have shown that in doped $III$-$V$ semiconductors, 
$Mn$-$As(N)$ is AFM polarized only for strong $p$-$d$ hybridization;
the $Mn$-$Mn$ spin interaction is double-exchange-like 
ferromagnetic coupling, addressing the strong ferromagnetism in diluted 
magnetic $III$-$V$ semiconductors. 
The consideration of the $p$-orbital degeneracy will lead to more 
reasonable polarized magnetic moment and polarization at $As$ or $N$ site.

\acknowledgements
   Authors appreciate the useful discussions with X. G. Gong, J. L. Wang 
and Q.-Q. Zheng. This work was supported by the  NSF of China, the BaiRen 
Project from the Chinese Academy of Sciences (CAS) and KJCX2-SW-W11.

\bibliography{apssamp}

\begin{thebibliography}{}
\bibitem{ref1}
 H. Ohno, D. Chiba, F. Matsukura, T. Omiya, E. Abe, T. Dietl, Y. Ohno and 
 K. Ohtanl, {\it Nature}, {\bf 408}, 944 (2000); F.Matsukura, H. Ohno, and 
 T. Dietl, {\it Handbook of Magnetic Materials}, {\bf 14},  p. 1-87 
 (2002).
\bibitem{ref2}
 M. J. Seong, S. H. Chun, H. M. Cheong, N. Samarth and A. Mascarenhas, 
 {\it Phys. Rev.} {\bf B 66}, 033202 (2002).
\bibitem{ref3}
 J. Szczytko, A. Twardowski, K. $\acute{S}$wiatek, M. Palczewska, M. Tanaka, 
 T. Hayashi and K. Ando, {\it Phys. Rev.} {\bf B 60}, 8304 (1999).
\bibitem{ref4}
 J. Okabayashi, T. Mizokawa, D. D. Sarma, A. Fujimori, T. Slupinski, A. 
 Oiwa and H. Munekata, {\it Phys. Rev.} {\bf B 65}, 161203 (2002).
\bibitem{ref5}
 P Kacman, {\it Semicond. Sci. Technol.} {\bf 16}, R25-R39 (2001).
 (2002).
\bibitem{ref6}
 H. Ohno, {\it Science}, {\bf 281}, 951 (1998)
\bibitem{ref7}
 F. Matsukura, H. Ohno, A. Shen and Y. Sugawara, {\it Phys. Rev.} {\bf B 57}, 
 R2037 (1998).
\bibitem{ref8}
 H. Ohno, {\it J. Magn. Magn. Mater.} {\bf 200}, 110 (1999).
\bibitem{ref9}
 K. Hirakawa, S. Katsumoto, T. Hayashi, Y. Hashimoto and Y. Iye, {\it Phys. 
Rev.} {\bf B 65}, 193312 (2002).
\bibitem{ref10}
 A. Kaminski and S. Das Sarma, {\it Phys. Rev. lett.} {\bf 88}, 247202 
(2002).
\bibitem{ref11}
 A. Zunger, {\it Solid State Phys.} {\bf 39}, 275 (1986).
\bibitem{ref12}
 A. Zunger and U. Lindefelt, {\it Phys. Rev.} {\bf B 27}, 1191 (1983).
\bibitem{ref13}
 C. Delerue, M. Lannoo and G. Allan, {\it Phys. Rev.} {\bf B 39}, 1669 
 (1989).
\bibitem{ref14}
 S. Sanvito, P. Ordej$\acute{o}$n and N. A. Hill, {\it Phys. Rev.}
 {\bf B 63}, 165206 (2001).
\bibitem{ref15}
 Yu-Jun Zhao, W. T. Geng, K. T. Park and A. J. Freeman, {\it Phys. Rev.} 
{\bf B 64}, 035207 (2001).
\bibitem{ref16}
 B. Beschoten, P. A. Crowell, I. Malajovich and D. D. Awschalom, {\it Phys. 
 Rev. lett.} {\bf 83}, 3073 (1999).
\bibitem{ref17}
 B. K. Rao and P. Jena, {\it Phys. Rev. lett.} {\bf 89}, 185504 (2002).
\bibitem{ref18}
 Shi-hao Wei, X. G. Gong, Gustavo M. Dalpian and Su-huai Wei, {\it Preprint}.
\bibitem{ref19}
 J. Okabayashi, A. Kimura, O. Rader, T. Mizokawa, A. Fujimori, T. Hayashi and 
 M. Tanaka,  {\it Phys. Rev.} {\bf B 58}, R4211 (1998).
\end{thebibliography}

\end{document}